\documentclass[twocolumn]{aastex62}
\usepackage{epsfig}
\usepackage{epstopdf}
\usepackage{sidecap}

\received{March 13, 2018}
\revised{July 30, 2018}
\accepted{August 20, 2018}
\submitjournal{PASP}

\shorttitle{Transiting Exoplanets with \textit{JWST}}
\shortauthors{Bean et al.}

\newcommand{\microns}{\micron}
\definecolor{orange}{rgb}{0.8,0.4,0}

\begin{document}

\title{The Transiting Exoplanet Community Early Release Science Program for \textit{JWST}}

\correspondingauthor{Jacob L. Bean}
\email{jbean@astro.uchicago.edu}

\author{Jacob L.\ Bean}
\affil{Department of Astronomy \& Astrophysics, University of Chicago, 5640 S.\ Ellis Avenue, Chicago, IL 60637, USA}

\author{Kevin B.\ Stevenson} 
\affil{Space Telescope Science Institute, 3700 San Martin Drive, Baltimore, MD 21218, USA}

\author{Natalie M.\ Batalha}
\affil{NASA Ames Research Center, Mail Stop 244-30, Moffett Field, CA 94035, USA}

\author{Zachory Berta-Thompson}
\affil{Department of Astrophysical and Planetary Sciences, University of Colorado, Boulder, CO 80309, USA}

\author{Laura Kreidberg}
\affil{Harvard-Smithsonian Center for Astrophysics, 60 Garden Street, Cambridge, MA 02138, USA}
\affil{Harvard Society of Fellows, 78 Mt.\ Auburn Street, Cambridge, MA 02138, USA}

\author{Nicolas Crouzet} 
\affil{Instituto de Astrof\'isica de Canarias, C. V\'ia L\'actea s/n, E-38205 La Laguna, Tenerife, Spain}
\affil{Universidad de La Laguna, Dept. de Astrof\'isica, E-38206 La Laguna, Tenerife, Spain}

\author{Bj\"{o}rn Benneke} 
\affil{D\'{e}partement de Physique, Universit\'{e} de Montr\'{e}al, Montreal, H3T 1J4, Canada}

\author{Michael R.\ Line}
\affil{School of Earth and Space Exploration, Arizona State University, PO Box 871404, Tempe, AZ 85287-1404, USA}

\author{David K.\ Sing}
\affil{Physics and Astronomy, College of Engineering, Mathematics and Physical Sciences, University of Exeter, EX4 4QL, UK}

\author{Hannah R.\ Wakeford}
\affil{Space Telescope Science Institute, 3700 San Martin Drive, Baltimore, MD 21218, USA}

% ...

\begin{abstract}
The \textit{James Webb Space Telescope} (\textit{JWST}) presents the opportunity to transform our understanding of planets and the origins of life by revealing the atmospheric compositions, structures, and dynamics of transiting exoplanets in unprecedented detail. However, the high-precision, time-series observations required for such investigations have unique technical challenges, and prior experience with \textit{Hubble}, \textit{Spitzer}, and other facilities indicates that there will be a steep learning curve when \textit{JWST} becomes operational. In this paper we describe the science objectives and detailed plans of the Transiting Exoplanet Community Early Release Science (ERS) Program, which is a recently approved program for \textit{JWST} observations early in Cycle 1. We also describe the simulations used to establish the program. The goal of this project, for which the obtained data will have no exclusive access period, is to accelerate the acquisition and diffusion of technical expertise for transiting exoplanet observations with \textit{JWST}, while also providing a compelling set of representative datasets that will enable immediate scientific breakthroughs. The Transiting Exoplanet Community ERS Program will exercise the time-series modes of all four \textit{JWST} instruments that have been identified as the consensus highest priorities, observe the full suite of transiting planet characterization geometries (transits, eclipses, and phase curves), and target planets with host stars that span an illustrative range of brightnesses. The observations in this program were defined through an inclusive and transparent process that had participation from \textit{JWST} instrument experts and international leaders in transiting exoplanet studies. The targets have been vetted with previous measurements, will be observable early in the mission, and have exceptional scientific merit. Community engagement in the project will be centered on a two-phase Data Challenge that culminates with the delivery of planetary spectra, time-series instrument performance reports, and open-source data analysis toolkits in time to inform the agenda for Cycle 2 of the \textit{JWST} mission.
\end{abstract}


\section{Introduction} \label{sec:intro}
All the stakeholders, from scientists to the general public, are eagerly awaiting the launch of \textit{JWST} with the anticipation that it will transform our understanding of planets and the origins of life. One of the primary ways that \textit{JWST} will make an impact on this topic is through observations of the atmospheres of transiting exoplanets. \textit{JWST} will be transformative in this area because of its capability for continuous, long duration observations and its dramatically larger collecting area, spectral coverage, and spectral resolution compared to existing space-based facilities \citep{beichman14}. This improvement, coupled with the myriad new exoplanets discovered by planet-searching programs like the upcoming \textit{TESS} mission \citep{sullivan15}, will give our community the opportunity to create the first comprehensive census of exoplanet atmospheres and push the characterization of individual planets to potentially habitable worlds \citep{cowan15}. Such a breakthrough will be an advance for the field of exoplanets comparable to that of the immensely successful \textit{Kepler} mission, thus fulfilling one of \textit{JWST}'s key promises.

\begin{figure*}
\begin{center}
\includegraphics[width=0.6\textwidth]{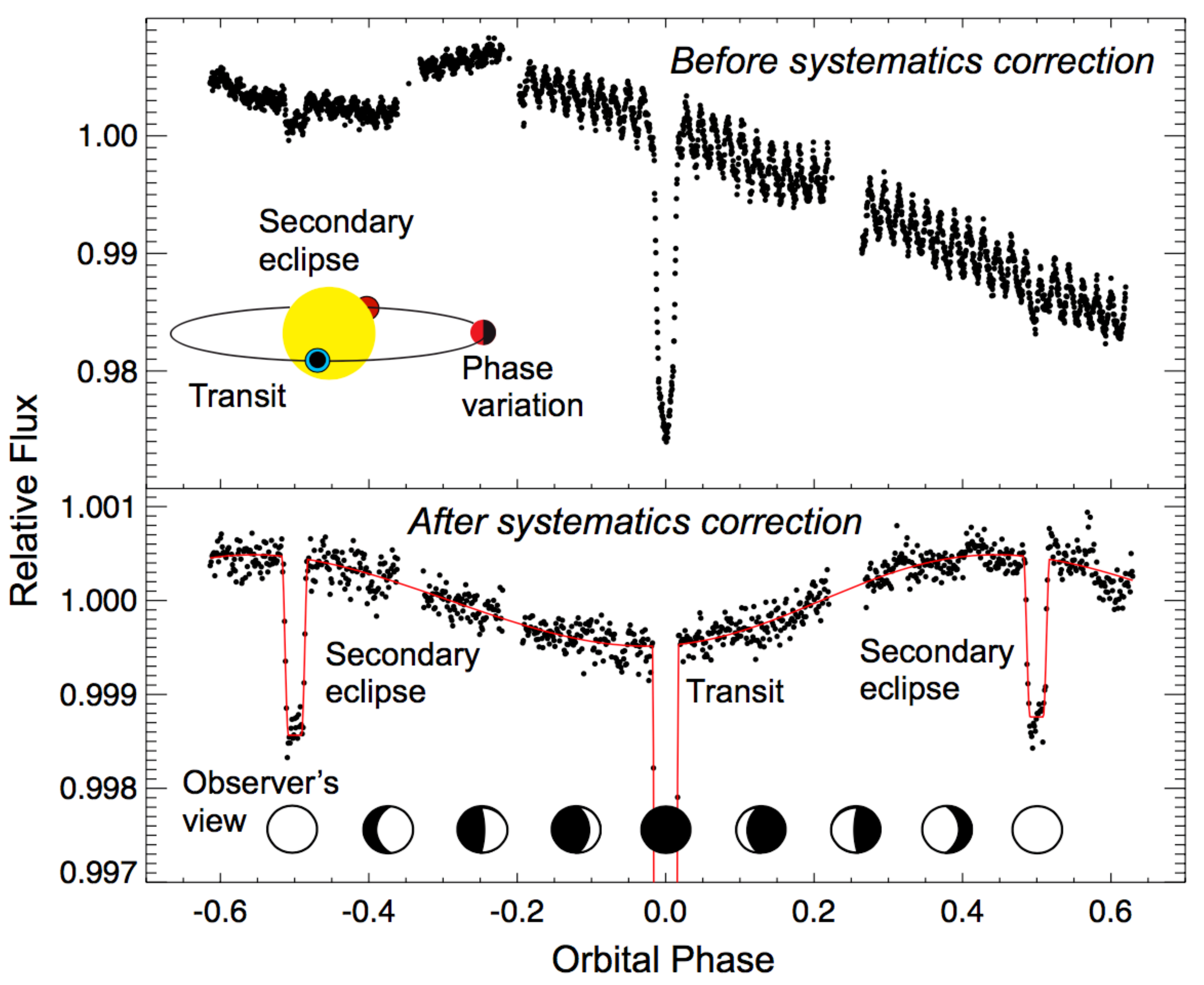}
\caption{\label{fig:systematics} Example \textit{Spitzer}/IRAC 4.5\,$\mu$m phase curve for the hot Jupiter HD\,189733b. The dominant instrumental signal in the raw photometry (periodic variations and long-term drift) is due to intrapixel sensitivity variations, coupled with an undersampled PSF and telescope pointing variations. Gaps in the coverage are due to spacecraft downlink breaks. We expect qualitatively similar instrument systematics for \textit{JWST} time-series observations. Our ERS program is designed to deliver representative datasets to the community to accelerate the development of strategies for modeling and removing these effects. Figure adapted from \citet{knutson12}.}
\end{center}
\end{figure*}

Despite the intense interest in observations of transiting planets and the substantial amount of ground testing that has been done, the actual performance of \textit{JWST} for the ultra-stable time-series spectrophotometry required for these studies remains to be characterized. The science agenda of the transiting exoplanet community requires measuring fractional changes in stellar spectra to a precision of up to $10^{-5}$ (10\,ppm). This value is consistent with \textit{JWST}'s expected photon-limited noise for bright targets, but it is much more precise than the limit to which the instruments have been tested on the ground. For example, the broadband lamps used for time-series observations in Cryovac Test 3 were stable to only the $10^{-2}$ level \citep{giardino.2017.moeta3eaiase}. Tiny effects that remain to be characterized like small image motion coupled to subtly non-uniform pixels or finite slits can introduce instrumental systematics that dominate the noise budget for precise time-series data.

Determining how to obtain robust results using general-purpose facilities has been the main challenge in the field of transiting exoplanet atmospheres since the first successful observations more than 15 years ago \citep{charbonneau02}. For example, it is well established that standard pipeline-processed data for transiting exoplanets from \textit{Hubble} and \textit{Spitzer} exhibit instrument-specific systematic noise that is larger than the photon-dominated noise and the sought-after astrophysical signals \citep[e.g.,][see Figure ~\ref{fig:systematics}]{brown01, charbonneau05, pont07, swain08, berta12}. It is anticipated that \textit{JWST} time-series data will have similar but also particular systematics that must be corrected to pursue transiting exoplanet science \citep[for a review of the instrument systematics affecting high precision time-series observations see][]{beichman14}. 

The community has recovered near photon-limited performance from many existing instruments \citep[e.g., better than 20\,ppm precision has been obtained using {\it Hubble}/WFC3,][]{line16a}, and there is general consensus on the results from the most popular instruments like \textit{Spitzer}/IRAC and \textit{Hubble}/WFC3 \citep[e.g.,][]{ingalls16,wakeford16}. However, it took years of work to establish the best observational and data analysis strategies for these facilities. Since \textit{JWST} will have a short (relative to \textit{Hubble}) and finite lifetime, identifying the dominant systematics and developing solutions for deriving high-fidelity data products from the key observing modes early in the mission will be crucial to maximizing its impact on transiting exoplanet science.

While the technical challenges faced by the transiting exoplanet community are unique, the need to accelerate the entire astronomical and planetary science communities' knowledge of \textit{JWST} data and capabilities is general and has been anticipated by the Space Telescope Science Institute (STScI). This anticipation led to the creation of the Director's Discretionary Early Release Science (DD ERS) program, which is a mechanism for allocating Director's Discretionary Time for observations that will provide representative datasets to the \textit{JWST} user community soon after the commissioning of the observatory\footnote{More information on the DD ERS program can be found at this website: \url{https://jwst.stsci.edu/science-planning/calls-for-proposals-and-policy/early-release-science-program}.}. Proposals for the ERS program were due in August 2017 and the results of the selection were announced in November 2017\footnote{The list of selected DD ERS proposals can be found at this website: \url{https://jwst.stsci.edu/observing-programs/approved-programs}.}.

\begin{figure*}
\begin{center}
\includegraphics[width=0.9\textwidth]{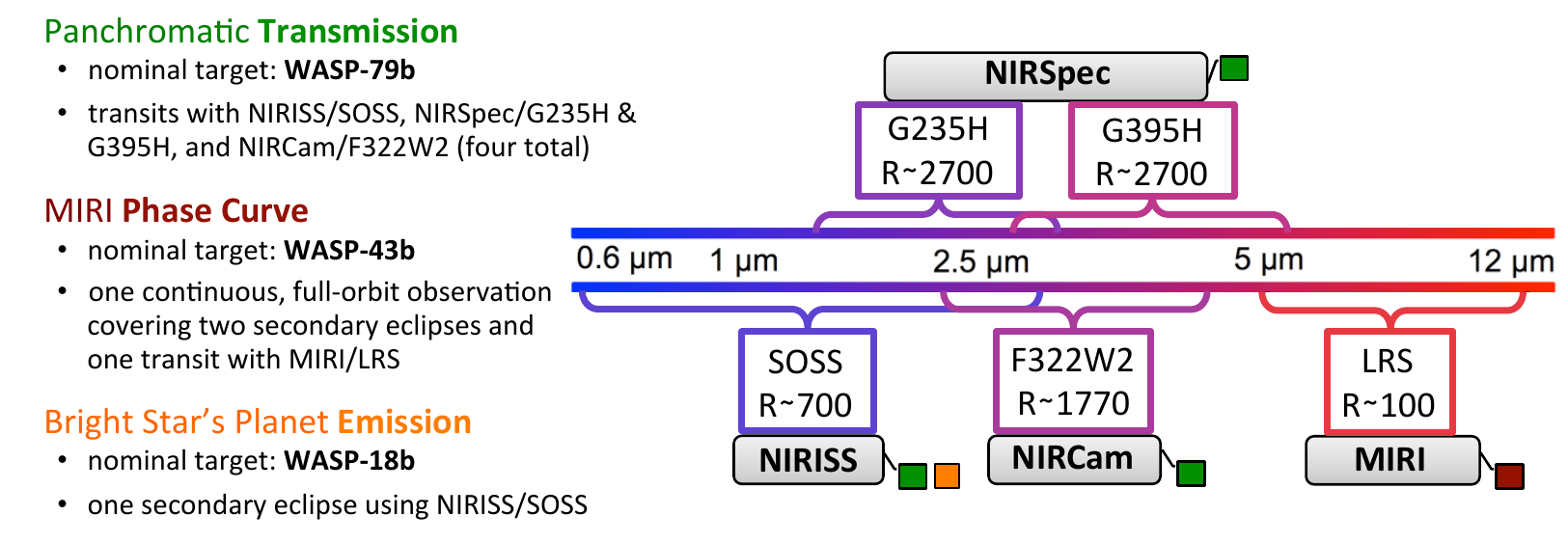}
\caption{\label{fig:roadmap} Summary of the three \textit{JWST} observing programs that comprise the Transiting Exoplanet Community ERS program. The schematic on the right indicates the wavelength coverage of the instrument modes that will be utilized. Note that the color coding on the text to the left corresponds to the instrument mode labels on the right.}
\end{center}
\end{figure*}

The transiting exoplanet community self-organized to respond to the DD ERS program and developed a successful proposal that was based on 22 months of inclusive and transparent work \citep[Proposal ID 1366, ][]{our_proposal}. The effort began at a workshop held at STScI in November 2015\footnote{\url{https://jwst.stsci.edu/news-events/events/events-area/stsci-events-listing-container/stsci-event-11?mwc=4}}. This early phase culminated in a community white paper that described the expected challenges of transiting exoplanet observations with \textit{JWST} and possible programs that could be proposed to elucidate the performance of the instruments \citep{Stevenson2016c}. In October 2016, an open call for participation in an ERS working group was extended to the transiting exoplanet community by NASA's Nexus for Exoplanet System Science (NExSS), a NASA research coordination network. Approximately 100 scientists contributed to the proposal planning. A second open workshop on transiting exoplanet science with \textit{JWST} was held at STScI in July 2017\footnote{\url{https://jwst.stsci.edu/news-events/events/events-area/stsci-events-listing-container/enabling-transiting-exoplanet-observations-with-jwst?mwc=4}}. The community reached consensus on the technical and scientific objectives, and a proposal went forward with 61 investigators and 43 collaborators. The community program has been allocated approximately 80 hours of \textit{JWST} time to perform three scientific and technical investigations that together comprise six separate observations.

The goal of this paper is to describe the plans and expectations of the Transiting Exoplanet Community ERS program. In \S\ref{sec:overview} we give an overview of the strategic objectives, observations, underlying philosophy, management structure, and timeline of the program. In \S\ref{sec:panchromatic}, \ref{sec:miri}, and \ref{sec:bright} we describe the details of the three investigations that make up our observing program. A key element of our community project is a comprehensive plan to quickly assess and disseminate our understanding of the performance of the instruments. This plan is presented in \S\ref{sec:analysis}. We conclude in \S\ref{sec:conclusion} with a look at the path ahead.

\section{Program Overview} 
\label{sec:overview}

\subsection{Objectives}
The goal of our \textit{JWST} ERS program is to deliver a substantial advance in the transiting exoplanet community's technical and scientific knowledge in advance of Cycle 2. We have three strategic objectives that constitute this goal:
\begin{itemize}
\item \textbf{Determine the spectrophotometric time-series performance of the key instrument modes on timescales relevant to transits for a representative range of target star brightnesses.} We will perform a quantitative assessment of the obtained data quality to inform future instrument and integration/exposure time selections, and shed light on the feasibility of characterizing potentially habitable exoplanets.
\item \textbf{Jump-start the process of developing remediation strategies for instrument-specific systematic noise.} We will develop open-source data analysis codes that will provide a foundation for establishing the best practices in removing systematics.
\item \textbf{Provide the community a comprehensive suite of transiting exoplanet data to fully demonstrate \textit{JWST}'s scientific capabilities in this area.} We will leverage these datasets to engage the community in an open Data Challenge that will deepen our understanding and accelerate the diffusion of this knowledge.
\end{itemize}

\subsection{Observations}
As mentioned above, our approved program is to perform three scientific and technical investigations that together comprise six separate observations. These three investigations are referred to as the Panchromatic Transmission, MIRI Phase Curve, and Bright Star Programs, and the observations utilize five of \textit{JWST}'s spectroscopic modes. The Panchromatic Transmission Program involves four transit observations of a single target with a total of three instruments: NIRISS (one observation with the SOSS mode), NIRSpec (one observation each in the G235H and G395H modes), and NIRCam (one observation in the F322W2 mode). The MIRI Phase Curve Program involves a full orbit phase curve of a planet using MIRI in the LRS slitless mode. The Bright Star Program is focused on a secondary eclipse observation of a planet orbiting a bright star using NIRISS in the SOSS mode. See Figure~\ref{fig:roadmap} for a summary of our observing plan. More details of the observing programs are given in the following sections.

The community ERS program was developed to be both comprehensive and efficient. Beyond the two workshops described above, the community process involved open email, message board, and telecon discussions, and democratic voting was used to make decisions in this framework. The resulting program is comprehensive in the sense that it takes advantage of the three transiting planet geometries (primary transit, secondary eclipse, and phase curves), covers all of the atmospheric observables (compositions, thermal structures, and dynamics on both the day- and night-sides), and utilizes all four \textit{JWST} instruments.

The program is also efficient because it includes only the highest priority observations that were identified from the community process. Other possible observations, while interesting, were judged lower priority given the ERS goals and thus were dropped from this self-limiting program. \added{For example, repeated observations of the same target with the same setup on the near-infrared (NIR) instruments was judged lower priority because we are already observing the same target with all three of the NIR instruments and four unique settings. By requiring a consistent transit shape (modulo the limb darkening) for each observation the cross comparison of these datasets will enable much the same test as repeated observations with the same setup.

Another possible observation that was judged as not an essential part of the program by the community was a phase curve with one of the NIR instruments. The MIRI phase curve was considered a higher priority because the long wavelength coverage of this instrument is more suitable for thermal emission measurements of a wider range of objects. Being limited to wavelengths $<$\,5\,$\mu$m means that the NIR instruments are only sensitive to the thermal emission of warm to hot objects. For example, GJ\,436b (T$_{\mathrm{day}}$\,$\approx$\,700\,K) is the coolest exoplanet with detected thermal emission in \textit{Spitzer}/IRAC's short wavelength bands \citep{deming07,demory07,stevenson10,lanotte14,morley17b}. The MIRI phase curve was seen as a test of the observatory-level issues with long-duration stares that would be relevant for all the instruments. Thus, the results of this observation in combination with the results of the other instrument-specific observations would mostly tell us what the community needs to know to plan future phase curve observations with any \textit{JWST} instrument.}

The strategy for all of our \textit{JWST} ERS observations is based on the best practices identified from nearly 20 years of space-based transit observations with \textit{Hubble}, \textit{Spitzer}, \textit{MOST}, \textit{CoRoT}, \textit{EPOXI}, and \textit{Kepler} \citep{beichman14}. This strategy includes 30 minutes of ``burn in'' (or settling) time for the telescope and infrared detectors to stabilize following a slew to a new target. We also allocate four hours of observations that, nominally, would be split evenly before and after transit/eclipse events to establish the baseline and characterize the instrument systematics. Observation start windows are all one hour in duration \added{to avoid the one hour ``tax'' that is imposed on observations with tighter start time constraints.} Therefore, the actual amount of baseline before ingress will be in the range of 2 -- 3 hours (including burn in).

For the phase curve program we will observe the full orbit of the targeted planet and begin and end with complete coverage of a secondary eclipse, again to enable accurate measurements. By taking this conservative strategy for the first \textit{JWST} observations \added{\citep[as compared to capturing just a fraction of the planet's orbit, e.g.][]{knutson.2007.dcep1}}, future studies of transiting planets will have confidence in the adopted approach, either justifying continued use of this strategy or reducing the amount of time needed to model and remove the instrument systematics. The technical details for the observations, including the output of STScI's proposal preparation and submission tool (called the Astronomer's Proposal Tool, or APT), are in the public domain\footnote{\url{https://jwst.stsci.edu/observing-programs/program-information?id=1366}}.

\subsection{Targets}
Following the criteria detailed by \cite{Stevenson2016c}, our targets were selected to: (1) have known, large signals from previous \textit{Hubble} and/or \textit{Spitzer} observations \citep[e.g., WASP-101b was recently found to be cloudy and thus was dropped as a possible target for the Panchromatic Transmission Program,][]{wakeford2017a}; (2) be observable early in the mission; (3) orbit quiet host stars; and (4) not be in conflict with Guaranteed Time Observations programs. We also endeavored, where possible, to choose targets that have long visibility windows (high ecliptic latitudes) and short visit durations (either transit, eclipse, or phase curve). Therefore, the selected observations strike a balance between maximal observing flexibility and minimal observing time required to obtain clear results.  These judicious observations will set the stage for more ambitious programs in later cycles.

\begin{figure}
\begin{center}
\includegraphics[width=0.45\textwidth]{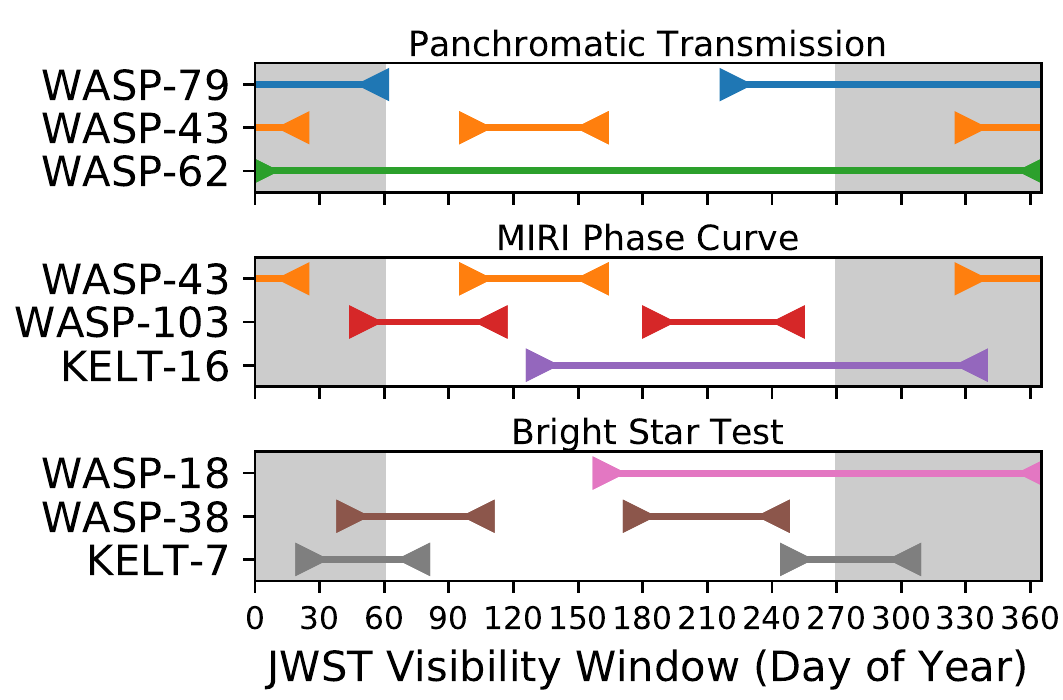}
\caption{\label{fig:visibility} \textit{JWST} visibility windows for the DD ERS program targets. The grey boxes indicate the first five months of science operations assuming a March 30, 2021 launch and six months of post-launch commissioning. The topmost targets for each program are the nominal targets assuming no further changes to the schedule.}
\end{center}
\end{figure}


Schedulability is a key issue in the ERS program because the targets will need to be observed very early in Cycle 1 to ensure that the data are available to the community in time to guide the design of observing programs for Cycle 2. The schedulability of our targets is summarized in Figure~\ref{fig:visibility}. We originally proposed a set of targets that would be suitable for observation in the first five months of science operations, assuming an October 2018 launch followed by a six month commissioning phase. \added{However, slips in the launch date (currently March 2021)} required us to switch our primary target for the Panchromatic Transmission Program from WASP-39b to WASP-79b. The nominal targets for the other two programs are unchanged at this point (see Table \ref{tab:planets}). The change in target has required an additional two hours of observing time due to the longer transit duration of WASP-79b as compared to WASP-39b (80.4 hours are now required and have been allocated for the program whereas 78.1 hours were originally requested). In addition to the primary targets, we have two backup targets for each program to ensure that the program can be executed early in Cycle 1 if there are additional changes to the \textit{JWST} schedule.

\added{Given the most recent slip in the {\em JWST} launch date and the successful launch of {\em TESS}, other high-ecliptic-latitude targets might be identified prior to finalizing the ERS target list in early 2019.  Under such a scenario, a significant amount of effort and resources would be required to measure the planet candidate's mass through radial velocity observations and perform atmospheric reconnaissance with WFC3.  Since there is no single ``best'' target for these programs and no guarantee that this hypothetical target will meet all of the criteria detailed by \cite{Stevenson2016c}, it remains likely that the final target list will contain planets that have already been discovered and are well characterized.}

\begin{figure*}
\begin{center}
\includegraphics[width=0.95\textwidth]{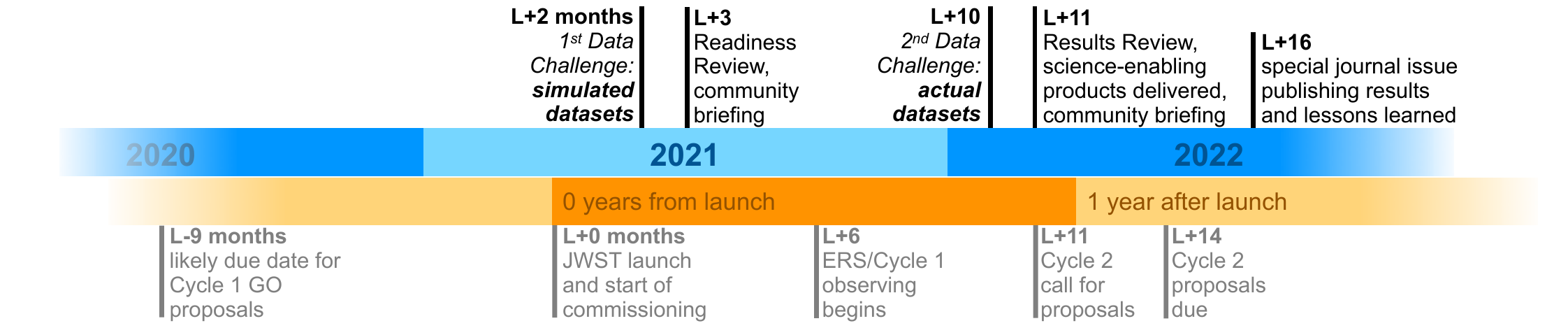}
\caption{\label{fig:timeline} Timeline of key \textit{JWST} mission and ERS program milestones, assuming \added{a March 30, 2021} launch.}
\end{center}
\end{figure*}

\subsection{Community and Management}
The management structure of the Transiting Exoplanet Community ERS program is based on best practices from the \textit{Kepler} mission and is designed so that our large, distributed, and diverse team can deliver products on time and on schedule. The PI and co-PIs were peer-elected. The team is organized into working groups for the three science programs and one for the Data Challenge (see \S\ref{sec:analysis}), each with its own leaders. A Science Council advises the project leadership and provides mediation for conflict resolution. Collectively, the team offers diversity of expertise (58\% observers and 33\% theorists, with 9\% self-identified as ``other'', which is a category that includes instrumentalists and administrators) and geography (54\% US, 46\% EU \& Canada), with members from research centers, research universities, and smaller undergraduate-focused institutions. Gender balance is 23\% women across the team (near the national average for astronomy), and 44\% women at leadership levels. A timeline for the project is given in Figure~\ref{fig:timeline}. \added{More information on the organization, policies, and progress of the project can be found on the program's website\footnote{\url{https://ers-transit.github.io}}.}

\begin{table*}[t]
\centering
\caption{\label{tab:planets} 
Properties of the nominal targets for the ERS program}
\begin{tabular}{ccccccccc}
    \hline
    \hline      
    Planet Name & Period    & Mass      & Radius    &T$_{\mathrm{eq}}$ & log\,g  & Tr. Depth & Tr. Dur.  & Reference    \\
                & (Days)    & (M$_{\mathrm{J}}$) & (R$_{\mathrm{J}}$) & (K)       & (cm\,s$^{-2}$)     & (\%)      & (Hours)   &              \\
    \hline
    WASP-79b    & 3.662     &  0.90     & 2.09      & 1760      & 2.88      & 1.3      & 3.99      & \citet{smalley12}   \\
    WASP-43b    & 0.813     &  2.03     & 1.04      & 1440      & 3.67      & 2.5      & 1.21      & \citet{gillon12}    \\
    WASP-18b    & 0.941     &  10.5     & 1.20      & 2400      & 4.26      & 1.2      & 2.18      & \citet{maxted13}     \\
    \hline
\end{tabular}
\end{table*}

\begin{table*}[t]
\centering
\caption{\label{tab:stars} 
Properties of the host stars for the nominal ERS targets}
\begin{tabular}{ccccccccccc}
    \hline
    \hline      
    Star Name   & Mass      & Radius    &T$_{\mathrm{eff}}$& log\,g  & Metallicity   & R.A.          & Dec.          & J-Band    & Reference             \\
                & (M$_{\odot}$) & (R$_{\odot}$) & (K)       & (cm\,s$^{-2}$)     & (Fe/H)        & (HH:MM:SS.ss) & (DD:MM:SS.s)  & (mag)     &                       \\
    \hline
    WASP-79     & 1.38      & 1.53      & 6600      & 4.20      &  0.03         & 04:25:29.02   & -30:36:01.5   &  9.3      & \citet{smalley12}   \\
    WASP-43     & 0.72     & 0.67     & 4520      & 4.65     & -0.01         & 10:19:38.01   & -09:48:22.6   & 10.0      & \citet{gillon12}    \\
    WASP-18     & 1.30     & 1.26     & 6400      & 4.35     &  +0.10         & 01:37:25.01   & -45:40:40.6   &  8.4      & \citet{maxted13}     \\
    \hline
\end{tabular}
\end{table*}

\section{Panchromatic Transmission Program} \label{sec:panchromatic}

\subsection{Scientific motivation: determining planetary nature and origins}
Atmospheric compositions are fundamental to our understanding of the nature and origins of planets. For example, the enhanced metallicities of the primary atmospheres of giant planets relative to their host stars' abundances are a tracer of these planets' formation histories \citep[e.g.,][]{owen99}. On the other hand, the compositions of the secondary atmospheres of terrestrial planets are a record of atmospheric evolution due to escape, geophysical, and, perhaps, biological processes \citep[e.g.,][]{meadows10}. For all planets, measuring the atmospheric composition is crucial to assessing and understanding planetary climate, including habitability.

Unfortunately, existing facilities give a very incomplete picture of transiting exoplanet atmospheric compositions. For example, the current best abundance measurements come from \textit{Hubble}/WFC3 ($\lambda$\,=\,0.8 -- 1.7\,$\mu$m, R\,$<$\,70), which primarily samples water \citep[e.g,][]{deming13}. The interpretation of such limited data is highly degenerate, and thus all of the results on this topic to date are dependent on substantial assumptions about the chemistries, elemental abundance ratios, aerosol properties, thermal structures, and homogeneity of the planets' atmospheres \citep[e.g.,][]{kreidberg.2014.pwamjw,wakeford.2017.hnewwhea}.

\textit{JWST} will be transformational for determining exoplanet compositions because it will have access to features from a much wider range of chemical species than existing facilities, and it will deliver data with the quality needed to break modeling degeneracies - even for cloudy planets \citep[e.g.,][]{benneke2012,line12,griffith14,barstow15,greene.2016.cteawj,rocchetto16,howe2017,batalha2017,chapman2017,molliere17}. This will enable astronomers to, for the first time, obtain a more complete chemical inventory of exoplanet atmospheres with fewer model assumptions, and thus fully capitalize on their potential for constraining planetary nature and origins.

\subsection{Technical motivation: exercising the NIR instruments}
Taking advantage of \textit{JWST}'s potential for transformational composition measurements will typically require multi-instrument observations to stitch together the needed wavelength coverage. Therefore, an exact understanding of, and a strategy for dealing with, the inevitable systematics will be crucial. As part of our ERS program we will obtain a panchromatic NIR (0.6 -- 5.2\,$\mu$m) transmission spectrum of a single object to demonstrate \textit{JWST}'s ability to obtain precise composition measurements for transiting planets and to exercise the instrument modes that will likely be the workhorses for observations of planets ranging from hot giants to temperate terrestrials. 

The Panchromatic Transmission Program has been designed to include the necessary wavelength coverage to cross-compare and validate the three NIR instruments, and thus establish the best strategy for obtaining transit spectroscopy measurements in future cycles. This program will test for agreement across four different modes with overlapping wavelengths (NIRISS/SOSS, NIRSpec/G235H, NIRCam/F322W2, NIRSpec/G395H), providing independent cross-validation of each mode. These three instruments probe a critical wavelength range that is dominated by the strongest bands of fundamental chemical species that have not been detected and/or precisely constrained before (e.g., CO, CO$_{2}$, CH$_{4}$, H$_{2}$S, HCN, and NH$_{3}$). Our technical goal is to characterize the NIR instruments at the precision necessary for exoplanet atmosphere studies, which requires a deep understanding of their time-series systematics.

\begin{figure*}
\begin{center}
\includegraphics[width=0.8\linewidth]{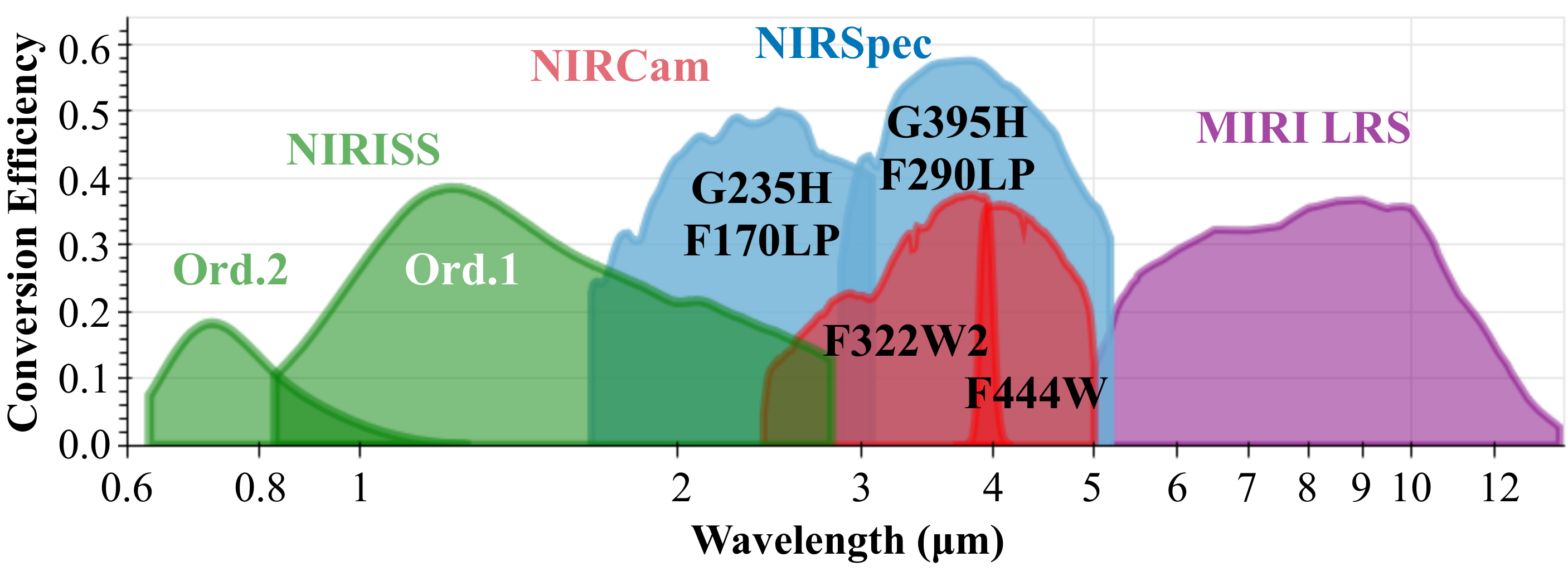}
\caption{\label{fig:throughputs}
Photon-to-electron conversion efficiencies for common NIRISS, NIRCam, NIRSpec, and MIRI instrument modes.  The conversion efficiency is the combined throughput from instrument optics, detector efficiency, quantum efficiency, and filter throughput (if applicable).  Multiple instruments and observing modes are required to obtain full (0.6 -- 12~{\microns}) wavelength coverage.  This program will test all of the instrument modes shown here, except for NIRCam's F444W mode.}
\end{center}
\end{figure*}

Obtaining data from all three NIR instruments on a single planet will enable the community to critically evaluate and compare the systematics associated with each mode. By comparing the overlapping wavelength regions using multiple instruments (the wavelength range 1.6 -- 4.0\,$\mu$m will have redundant coverage from at least two different instruments, see Figures~\ref{fig:roadmap} and \ref{fig:throughputs}), we will confirm the accuracy of our systematics corrections, validate the strength of common atmospheric features using multiple instruments, and determine the span of reliable wavelengths when stitching together spectra from multiple modes. This verification is especially critical in the 2 -- 3\,$\mu$m region, where the first and second orders of the NIRISS/SOSS mode overlap on the detector. This region is also covered by NIRCam/F322W and NIRSpec/G235H. Of these two modes, NIRSpec/G235H has a higher throughput and complete coverage of the overlapping orders, while NIRCam/F322W can observe significantly brighter stars.

\subsection{Planned observations}
\label{sec:planned_trans}
The nominal target for the Panchromatic Transmission Program is the hot Jupiter WASP-79b \citep{smalley12}, while the back-up targets are WASP-43b \citep{hellier11} and WASP-62b \citep{hellier12}.  WASP-79b has a transit duration of 3.75\,hours.  \added{We require $\sim$2 hours of baseline both before and after the transit to identify and correct the systematics plus 30 minutes to account for the uncertainty in the start time. Including overheads, we require 10.5 hours of telescope time per visit, or 42 hours for all four visits.  See Tables \ref{tab:planets} and \ref{tab:stars} for target system properties.  

Next, we describe the WASP-79b observations in detail with the caveat that these plans are still subject to change; therefore, the most up-to-date specifics for each observation can be found in the Observing Description section of the APT file.

The NIRISS/SOSS observations use the GR700XD grism (in combination with a clear filter) to obtain spectroscopy over 0.6 -- 2.8\,$\mu$m.  Since our primary target has a J-band magnitude of 9.3, we will utilize the ``nominal'' SUBSTRIP256 subarray, which is 256x2048 pixels, to acquire both 1$\sp{st}$ and 2$\sp{nd}$ orders simultaneously.  With 4 groups per integration (27.5 sec) and 1081 integrations per exposure, we will achieve a total exposure time of 8.25 hours at ~65\% of saturation.  The total science time is 6.6 hours, marking an 80\% observation efficiency (science time divided by total exposure time).  Target acquisition (TA) for SOSS is performed using the 64x64 subarray and F480M filter.  Using 11 groups and the SOSS Faint TA mode, we will achieve a signal-to-nois ratio (SNR) of $\sim$316 on our target (a minimum SNR of 20 is required for all TA).

The two NIRSpec observations are conducted in BOTS (Bright Object Time Series) mode, which requires the S1600A1 aperture with a fixed 1.6"x1.6" field of view.  Both exposures will use the SUB2048 subarray (2048x32 pixels) to record the full spectrum.  The first exposure will use the G235H+F170LP grating/filter combination (1.66 -- 3.17\,$\mu$m) with 13 groups per integration (12.6 sec) and 2352 integrations total.  The second exposure will use the G395H+F290LP combination (2.87 -- 5.27\,$\mu$m) with 26 groups per integration (24.3 sec) and 1220 integrations total.  Each 8.25-hour observation is designed to stay below 70\% of saturation and achieves an efficiency of $>$90\%.  Because our science target is too bright for target acquisition, we will utilize the Wide Aperture Target Acquisition (WATA) mode on a nearby, faint star (J=22.4) with the F110W filter and achieve a SNR of 21.

\begin{figure*}
\begin{center}
\includegraphics[width=1.0\linewidth]{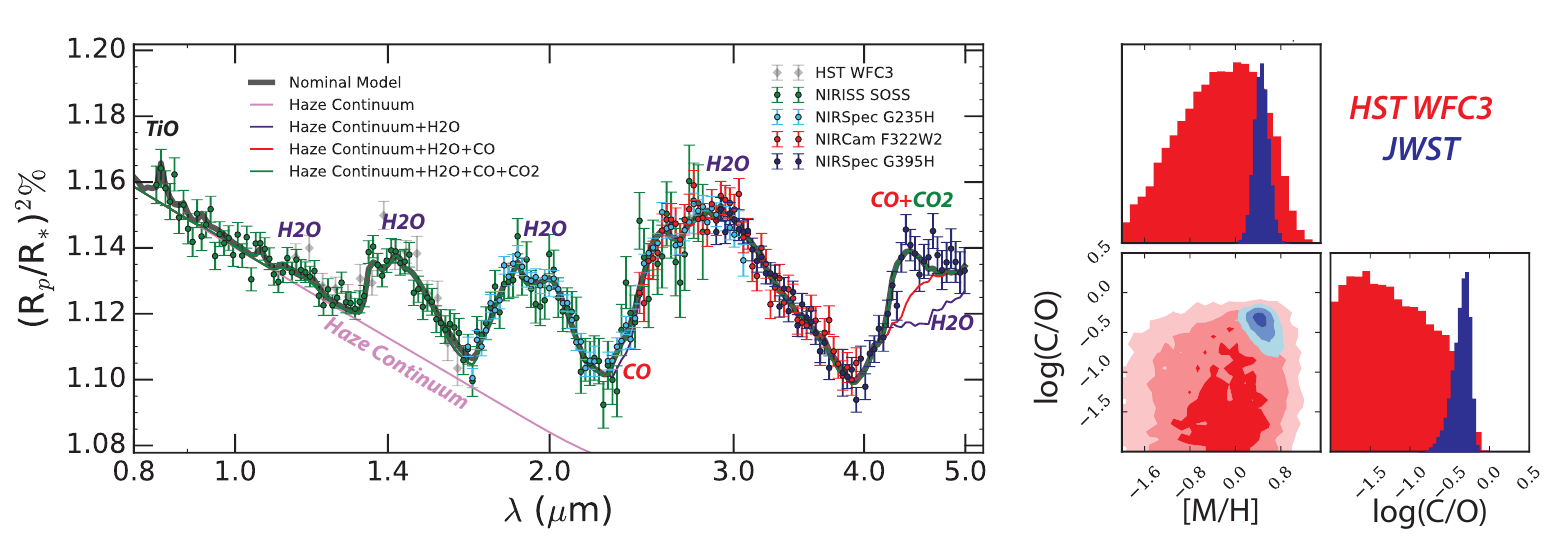}
\caption{\label{fig:w79}
Simulated data (left) and anticipated abundance constraints (right) from the Panchromatic Transmission Program observations of WASP-79b. Left: the points with error bars show the simulated observations from \texttt{PandExo} \citep{batalha.2017.pctteswj} for each mode (NIRISS SOSS-green, NIRSpec G235H-cyan, NIRCam F322W-red, NIRSpec G395H-blue) based on a nominal model fit (black) to the existing \textit{Hubble}/WFC3 data (gray). The contributions from the major opacity sources are also indicated. Right: anticipated constraints on the atmospheric metallicity and carbon-to-oxygen ratio (blue) compared with constraints from \textit{Hubble} (red).  These constraints are marginalized over aerosol properties, temperature-pressure profile, and an uncertain planetary reference radius.  The abundance constraints are improved by orders-of-magnitude over \textit{Hubble} due to the presence of multiple molecular features (e.g., \citealt{greene.2016.cteawj}). }
\end{center}
\end{figure*}

The NIRCam observation will use the grism time-series observing mode with the F322W2 filter (2.4 -- 4.0\,$\mu$m). In this mode, the Module A Grism R is used to disperse the target spectrum across detector pixel columns.  We will use the SUBGRISM256 subarray mode with a single output amplifier.  With 6 groups per integration (37.1 sec), we require 802 integrations for a 8.25 hour exposure.  The observation efficiency is 86\%.  We will position the target at one of the defined field points on the detector to obtain the full wavelength coverage within the subarray.  We will use 9 groups in the SUB32 subarray to perform target acquisition and achieve a SNR of $\sim$255 on our target.
}

\subsection{Expected results}
Figure~\ref{fig:w79} shows simulated data for the WASP-79b panchromatic transmission spectrum assuming photon-limited errors as estimated using the \texttt{PandExo} package \citep{batalha.2017.pctteswj}. We also show the results of a retrieval on this simulated spectrum using the \texttt{CHIMERA} code \citep{line14,line16b,batalha.2017.icasojnmtea} to predict the atmospheric constraints that will be obtained from these data. 
\added{The simulated data are based on an extrapolation of the best-fit model fit to existing \textit{HST}/WFC3 data, which indicate the presence of water absorption with a strong scattering haze slope toward the  optical (K. Showalter, et al. in prep). The best-fit model was retrieved using the \texttt{CHIMERA} code, which uses the nested Bayesian sampler \texttt{PyMultiNest} to derive the posterior distributions of the model parameters. The model is composed of multiple gas phase opacities, including the dominant absorbers (H$_2$O, CO, and CO$_2$) that inform the retrieved atmospheric metallicity ([M/H]) carbon-to-oxygen ratio (C/O). The abundance constraints based on the retrieval of the existing \textit{HST}/WFC3 data are shown in red on the right panel of Figure~\ref{fig:w79}. The best-fit model is then used with the \texttt{PandExo} package to simulate the intrinsic scatter in the measured transit depth based on \textit{JWST} instrument models, as well as the uncertainty in the measured data based on the system parameters (see Tables \ref{tab:planets} and \ref{tab:stars}) and the instrument characteristics (see \citealt{batalha.2017.pctteswj} for more details).  The resulting abundance constraints from the {\em JWST} retrieval are shown in blue on the right panel of Figure~\ref{fig:w79}.  It is evident from these results that a single transit from each mode is sufficient to characterize the atmosphere of this hot Jupiter.}

For this specific science investigation we focus on the key giant planet formation tracers metallicity and C/O \citep{lodders.2004.jfwmt,mousis.2009.dmmheejs, mousis.2012.nwdcjoa,madhusudhan.2011.cgpactisfc, madhusudhan.2014.tccjm,oberg.2011.espa,fortney.2013.fcalltp,helled.2014.mjwajlbifm,marboeuf.2014.fppvm,venturini.2016.pfweeipd,mordasini16,espinoza.2017.melartge}. Previously, only two transiting exoplanet metallicity measurements have accounted for both oxygen- and carbon-bearing species, and both of these assessments have substantial model dependencies \citep{Stevenson2017a,brogi17}. Furthermore, no observations have yielded robust and precise constraints on the C/O other than upper limits \citep[e.g.,][]{brogi14,brogi17,kreidberg15,kreidberg18a,benneke15,wakeford.2018.W39}. 

With large H$_2$O features already detected by \textit{Hubble}/WFC3 for our proposed target, all of the major molecular constituents expected for this planet (H$_2$O, CO, and CO$_2$) will be resolved at high S/N with \textit{JWST}. This will enable an empirical assessment of the metallicity and elemental abundance ratios by simply summing the retrieved abundances (e.g., the C/O can be computed by taking the ratio of the sums of the carbon- and oxygen-bearing species' abundances), thereby providing the community with its first comprehensive constraints on the composition of a transiting exoplanet. The simulated data suggest constraints on the metallicity and log(C/O) on the order of a factor of $\sim$2. This is a remarkable improvement over the \textit{Hubble} data alone, primarily driven by the presence of multiple species, \added{multiple absorption bands per species}, and higher SNR, a direct consequence of a larger telescope and broader wavelength coverage.   

In addition to revealing the abundance of the major expected molecules assuming a slightly metal-enhanced atmosphere in chemical equilibrium, the Panchromatic Transmission Program has the potential to reveal subtle features that could yield deeper insight to the properties of exoplanet atmospheres. This includes the detection of molecules that have not yet been clearly detected \citep[e.g.,][]{macdonald17}, the unique identification of aerosol species \citep[e.g.,][]{wakeford15,pinhas17}, definitive evidence of non-equilibrium chemistry \citep{stevenson10,moses11,line13,drummond18}, and the detection of non-uniform cloud coverage \citep[e.g.,][]{Fortney2010,line16b,kempton17}. Furthermore, by delivering a high SNR and high resolution transmission spectrum that covers many wavelengths for the first time, the Panchromatic Transmission Program has substantial potential for the discovery of unexpected phenomena.

\section{MIRI Phase Curve Program} \label{sec:miri}

\subsection{Scientific motivation: mapping climate, chemistry, and clouds}
In addition to being a record of origins and a diagnostic of planetary nature, atmospheres also govern planetary climate by mediating the balance between stellar irradiation, re-radiated flux, and a planet's intrinsic luminosity. Many of the known transiting exoplanets are highly irradiated, and their rotation rates are also strongly influenced by tidal forces. These factors give rise to faster winds, more dramatic temperature gradients, and larger spatial variations in chemistry and cloud cover compared to their Solar System counterparts \citep[e.g.,][]{showman02,menou03,fortney06,burrows10,lee16,lines18}. Therefore, the atmospheres of these objects are laboratories for planetary physics and chemistry in new regimes.

To truly understand the many physical phenomena at play in these atmospheres, we need to determine their \emph{three-dimensional} compositions and temperature structures. Such determinations are possible only with time-resolved spectroscopy over a planet's complete orbital revolution (i.e., a ``phase curve'') and during secondary eclipse ingress and egress \citep[i.e., using the technique of ``eclipse mapping'';][]{cowan17}. Eclipse mapping has so far only been used for one planet, HD~189733b, due to the limited sensitivity of existing facilities \citep{dewit12,majeau12}. On the other hand, phase-curve observations of close-in exoplanets have been a major focus of atmosphere characterization efforts to date, but these observations have also been held back by the limited capabilities of existing instruments in much the same way as described in Section \ref{sec:panchromatic}. \textit{Spitzer} has been the main facility used for both eclipse mapping and phase curves but can only perform photometric measurements \cite[e.g.,][]{knutson.2007.dcep1}, the interpretation of which is highly degenerate. \textit{Hubble}/WFC3 has been used in recent years to obtain spectroscopic phase curves \citep[e.g.,][]{Stevenson2014c,kreidberg18b}, but with limited spectral coverage and resolution, and also only for a handful of very short-period ($P$\,$<$\,1.5\,d) giant planets due to constraints on long-duration stares and poor sensitivity.

\textit{JWST} will yield a major advance in our understanding of planetary physics and chemistry because it will enable both phase curve and eclipse mapping observations to reveal the global composition and climate of a wide range of planets. Moreover, mid-infrared observations hold the promise of revealing the climate of terrestrial exoplanets for the first time through spectroscopic measurements at wavelengths where these planets emit most of their energy and the planet-to-star flux ratio is most favorable \citep{selsis2011,kreidberg.2016.pcapc,meadows16,morley17a}.

\subsection{Technical motivation: exercising MIRI and testing long-duration observations}
The MIRI Phase Curve Program will test the hour-to-hour stability of \textit{JWST} and MIRI/LRS \citep{Rieke2015a, Kendrew2015}. Phase-curve observations pose unique challenges that will not be tested with shorter transit- or eclipse-only observations. Ground-based performance evaluations are also inadequate to evaluate the stability of a bright source at the required level of a few tens of ppm on timescales of hours to days because they are limited by the stability of the source itself over such long durations. \added{MIRI is the most sensitive to thermal background of all the \textit{JWST} instruments and is the only one that is actively cooled, to 7~K}. Besides, the MIRI Si:As (arsenic-doped silicon) detectors are fundamentally different than the HgCdTe detectors used for \textit{JWST}'s NIR instruments \citep{Rieke2015b}. \added{Observing a very bright source for such a long duration is a unique operating regime for these detectors. 

This program will evaluate the following key points:
\begin{itemize}
\item High-gain antenna moves. They occur every 10,000\,s and may disrupt the pointing. By definition, the TSO (Time Series Observations) mode overrides the 10,000\,s exposure limit, and ten high-gain antenna moves should occur during the observation. We will search for increased noise and possible jumps in the lightcurves at the time of the moves.
\item Variations caused by the thermal background. We will correlate the lightcurve and background pixel variations with the spacecraft and instrument temperature telemetry data. 
\item Detector response drifts. We will investigate drifts in the detector response function over long timescales. Prior experience with {\em Hubble} and {\em Spitzer} \citep[\textit{e.g.}][]{Stevenson2014c, Stevenson2017a} has shown that standard spacecraft calibration data do not characterize such effects at the required precision. Separating out response drift from the signal is important for the phase curve analysis, but additionally this uniquely long MIRI exposure will allow us to identify any new drifts not previously seen in ground test data. The analysis will involve grouping the pixels of the SLITLESSPRISM subarray into flux bins, \textit{i.e.} pixels that show similar signal, and tracking the flux response around the average value for the entire observation. We will also seek to identify any changes in observing strategy that might mitigate their impact, such as the pre-flash technique implemented by {\em Spitzer} \citep{knutson.2009.149PH,ballard14}. 
\item Cosmic ray latency. The MIRI pipeline identifies cosmic rays, so we will use those flags to track the persistence decay from cosmic rays over time. Also, this very long data set with relatively long ramp lengths will allow us to test the cosmic ray hit rate. These results will be compared to the expectations \citep[\textit{e.g.}][]{Ressler2015}.
\item Pointing drifts combined with intra-pixel sensitivity variations and flat-field errors. We will correlate the lightcurve variations with the position of the spectral trace on the detector and evaluate their impact.
\item Reset Switch Charge Decay (RSCD), seen as a non-linear trend in the first frames of the ramp. We will fit this trend at the start of the ramp across every integration and measure its amplitude, duration, and stability over 30 hours. Current ground tests on shorter durations indicate that it is very predictable.
\end{itemize}

We also plan to obtain contemporaneous observations of the target with independent space- or ground-based facilities in order to monitor stellar variations and disentangle them from the planetary phase curve and long-term instrumental effects.} Finally, we will determine how the photometric precision improves when binning over multiple, independent transit durations. If we can demonstrate sufficient stability over $\sim24$ hours, future observations may be able to utilize shorter or non-continuous segments of data to measure exoplanet phase variations \citep{harrington06,krick16}.

\subsection{Planned observations}
We will observe a full-orbit phase curve (including two eclipses and one transit) of a very short-period hot Jupiter with MIRI LRS in slitless mode to make the first demonstration of mid-infrared (5 -- 12\,$\mu$m) phase-resolved spectroscopy. \added{To evaluate the long term stability of the instrument, the phase curve observation will be completed in a single visit: observing only parts of a phase curve would compromise the removal of expected systematics, and implementing multiple visits is inefficient due to significant overheads for time series observations.} The observation will start shortly before secondary eclipse and end shortly after the following secondary eclipse, including a single primary transit. This observing strategy will allow us to isolate the astrophysical signal from long-term instrument drifts by calibrating our results to the ``star only'' flux measured during both secondary eclipses that bracket the observation: \added{variations between both eclipses will reveal instrumental systematics, and matching both eclipses will allow us to anchor the flux baseline and isolate the phase curve signal, as illustrated in Figure \ref{fig:systematics}}. 

We focus on very short-period ($\leq 1$ day) hot Jupiters because the time required for the observation is relatively small (e.g., compared with a more typical hot Jupiter having P\,$\sim$\,3\,d), and they are highly irradiated and thus will give a large thermal emission signal. They are also the easiest to observe with current facilities and thus most have substantial existing data to complement and compare the new \textit{JWST} data against. The nominal target is \mbox{WASP-43b} \citep[$P$\,=\,0.81\,d,][see Table \ref{tab:planets} for target system properties]{hellier11} and the backup targets are \mbox{WASP-103b} \citep[$P$\,=\,0.93\,d,][]{gillon14} and \mbox{KELT-16b} \citep[$P$\,=\,0.97\,d,][]{oberst17}. \added{To the 19.5-hour period of WASP-43b, we add 1.2 hours to cover the duration of the second secondary eclipse and four hours of baseline to adequately measure the anticipated systematics while allowing for a one hour observation start window. The baseline is distributed as follows: 2.25${\pm}$0.5 hours before the first eclipse and 1.75${\pm}$0.5 hours after the second eclipse.  The distribution is asymmetric to leave ample time for the telescope to settle before the first eclipse measurement while still providing adequate flux baseline.
}

\added{The observations will use the MIRI/LRS SLITLESSPRISM mode without dithers. This yields 5 -- 12\,$\mu$m spectra with resolving power $\sim$100 (40 to 160 over the 5 -- 10\,$\mu$m range). This slitless mode is recommended for time-series observations, since minor pointing instabilities would otherwise result in a time-dependent systematic in the measured flux that is degenerate with the astrophysical signal. We used the \textit{JWST} Exposure Time Calculator (ETC) and the Astronomer's Proposal Tool (APT) to set up the observation: we will obtain a single 24.7-hour exposure containing 8,595 integrations of 65 groups each (one frame per group, 10.3\,sec per integration), leaving us well below the saturation limit of the detector ($\sim$60\%).  This observing strategy yields 24.3 hours of science time ($>$98\% observation efficiency) and, when including overheads, accounts for 29.6 hours of charged time.
The long duration of this exposure will provide an excellent opportunity to test persistence and detector stability over day-long timescales, understand the repeatability of high gain antenna moves (which occur every 10,000 seconds), and mitigate their impact on time-series observations. 

We will select the F1500W filter for target acquisition. This filter is adjacent to the LRS prism in the filter wheel; therefore, the star will not irradiate the detector through intermediate filters as the wheel rotates into position. Using 5 groups yields sufficient SNR ($\sim$475) to perform successful target acquisition without causing a bright persistent image.}

\subsection{Expected results}
WASP-43b is one of the best-characterized transiting exoplanets, but major questions with broad implications remain despite our best efforts with current facilities. Existing \textit{Hubble} and \textit{Spitzer} data show a day-night temperature contrast of at least 1000\,K, raising the possibility of striking variations in atmospheric chemistry and cloud coverage with longitude \citep{Stevenson2014c,Stevenson2017a,kataria15,mendonca18}. 

\begin{figure}
\begin{center}
\includegraphics[width=0.45\textwidth]{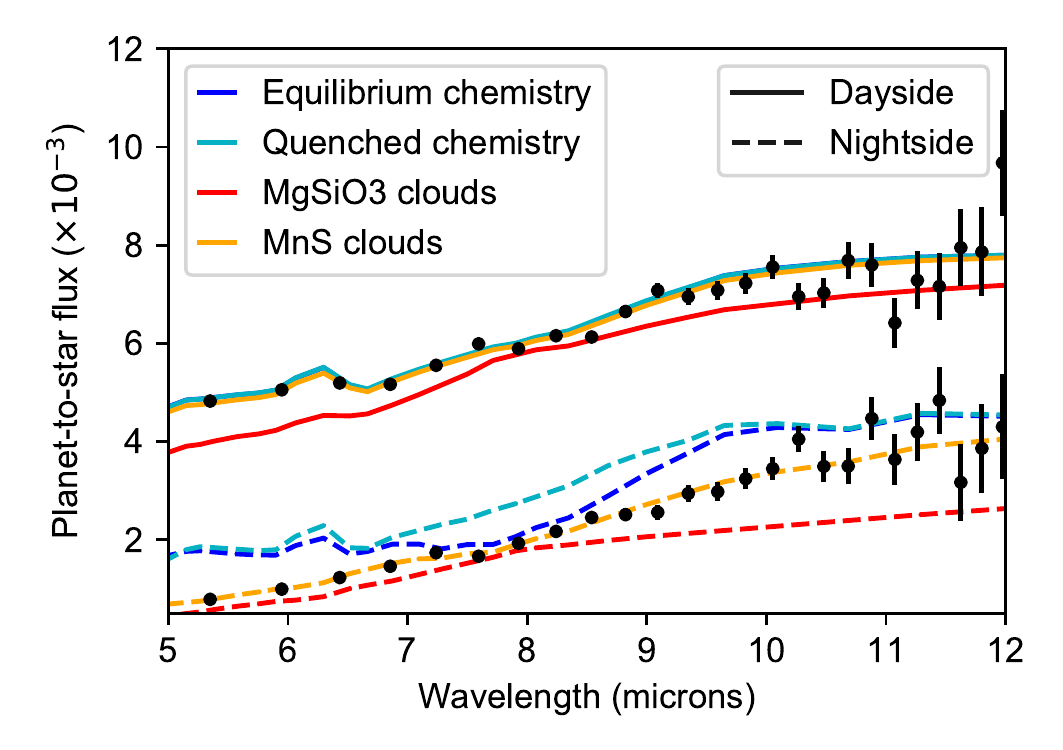}
\caption{\label{fig:w43_daynight_spectra} Simulated dayside and nightside thermal emission spectra for WASP-43b from a range of GCMs \citep[solid and dashed lines, respectively;][]{parmentier16}, compared to simulated MIRI observations (points with 1\,$\sigma$ uncertainties).}
\end{center}
\end{figure}

\begin{figure}
\begin{center}
\includegraphics[width=0.45\textwidth]{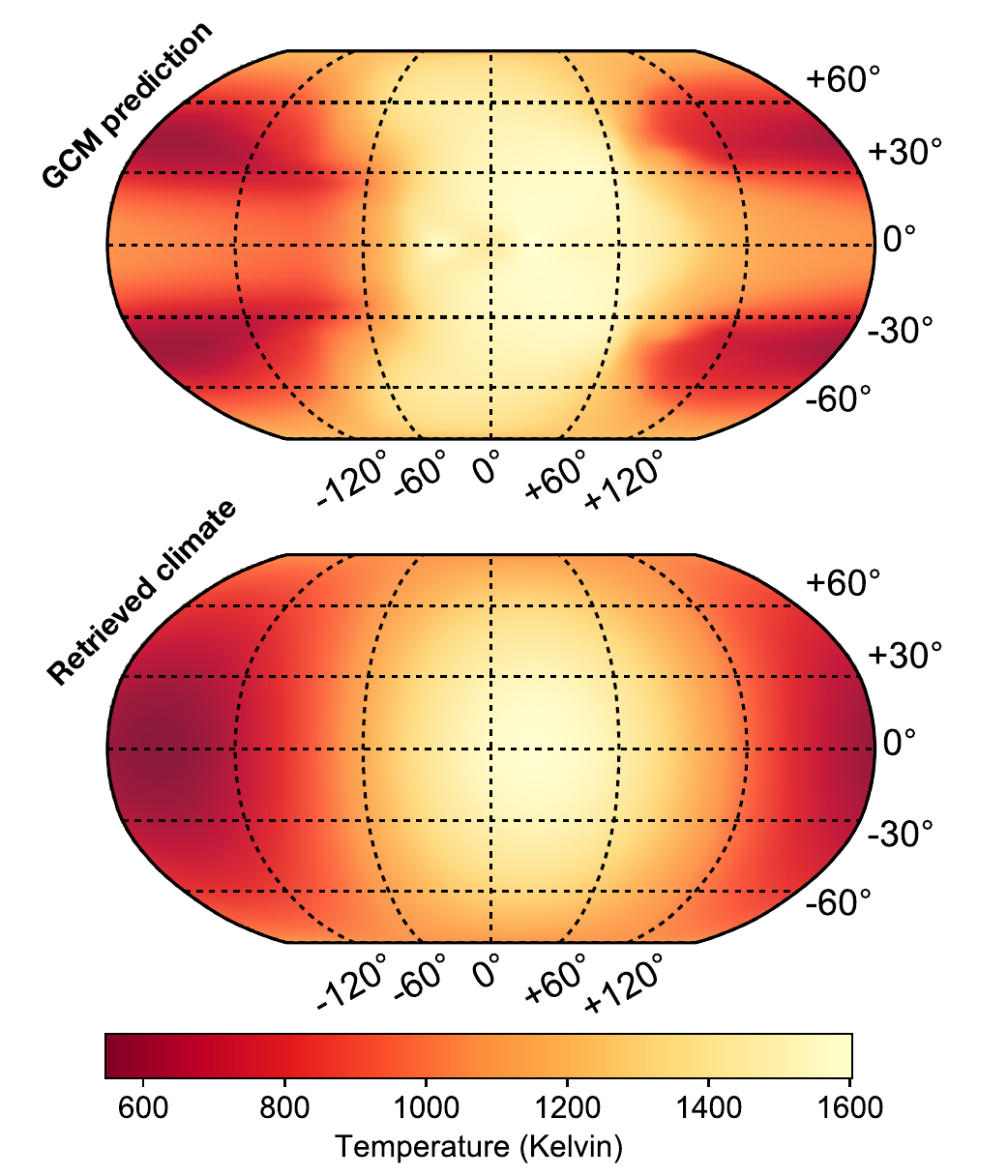}
\caption{\label{fig:w43_climate} Predicted temperature map for WASP-43b from the nominal cloud-free, solar composition GCM \citep[top panel;][]{kataria15} compared to a spherical harmonic map generated from the best fit to the simulated phase curve observations (bottom panel). }
\end{center}
\end{figure}

\added{To assess how our observations will address these questions and characterize the planet's atmosphere, we developed a theoretical framework to interpret the data that includes several models with a range of complexity. First, we used the 2D radiative/convective/advective equilibrium model of~\citet{tremblin:2017aa} to determine the thermal structure at the equator of the planet. The longitudinally varying temperature was then used to calculate the expected chemical state of the atmosphere considering vertical mixing ~\citep{venot:2012aa} and horizontal mixing (using an adaptation of the~\citet{moses2011} model following the method of~\citet{agundez2014aa}). We predict that the dayside photosphere should be close to thermochemical equilibrium, but the nightside of WASP-43b is expected to be quenched with a CO/CH$_{4}$ ratio close to 0.001.

We then used the 3D Global circulation model SPARC/MITgcm \citep{showman2009} to model the three-dimensional thermal structure of WASP-43b assuming either chemical equilibrium or a quenched abundance of CO/CH$_{4}$ of 0.001, inspired by the 2D chemical models. Both cloud-free and models with passive clouds were run. The outgoing spectrum at each phase were computed following~\citet{Parmentier2016}. All the models have a similar dayside spectrum but can be very well differentiated through the nightside spectrum. Multiple combinations of cloud composition (e.g. MnS of MgSiO$_3$) and physical properties (e.g. particle size) allow for a much better fit of current nightside observations ~\citep[see also][]{kataria2015,mendonca18}. More details and discussions about this modelling work will be presented in a forthcoming paper (Venot et al. in prep).

We used the output of the 3D GCM models to generate synthetic phase-resolved emission spectra, shown in Figure\,\ref{fig:w43_daynight_spectra}. We then simulated MIRI observations with the Pandexo tool \citep{batalha.2017.pctteswj} to estimate the measurement precision for the wavelength-dependent eclipse depths. We also generated phase curves from the GCM using the \texttt{SPIDERMAN} package \citep{louden18}. These phase curves include the eclipse mapping signature, which encodes 2D information about the planet's surface brightness distribution \citep{dewit12,majeau12}. We then fit the phase curve with a 2D climate map composed of spherical harmonics. Figure\,\ref{fig:w43_climate} shows the GCM temperature map compared to the \texttt{SPIDERMAN} retrieval.}

\added{Based on these simulations, we estimate that the WASP-43b phase curve will:} \\
\textit{Characterize the global climate.} We will measure the temperature-pressure profile to 30-Kelvin precision in 20 orbital phase bins (Figure~\ref{fig:w43_climate}). MIRI data sample the peak of the planet's emission on the nightside, enabling us to close the planet's energy budget and measure the Bond albedo to better than 1\%. We will determine the fraction of energy incident on the dayside that is transported to the nightside, and estimate the longitude of peak brightness to within one degree and its variations as a function of wavelength to set tight limits on the efficiency of advection relative to re-radiation at a wide range of pressures. In addition, we will use the eclipse mapping technique to map the dayside brightness temperature as a function of latitude and longitude \citep[e.g.,][]{dewit12}. \\
\textit{Constrain the dominant cloud species and particle size.} As illustrated in Figure~\ref{fig:w43_daynight_spectra}, varying cloud compositions are expected to produce observable differences, particularly on the cold nightside ($<$\,1000\,K) where MIRI is most sensitive. With only shorter wavelength data, the effect of clouds is degenerate with the effect of drag and disequilibrium chemistry. MIRI observations are necessary to break these degeneracies. \\
\textit{Retrieve the abundance of major absorbing species} (H$_2$O, CH$_4$, \textit{and} CO). These measurements will determine the overall metallicity and carbon/oxygen ratio in the atmosphere to shed light on the planet's origins \citep[e.g.,][]{mordasini16}. We will also determine how the abundances change with longitude to constrain the effects of transport-induced quenching \citep[][see Figure~\ref{fig:w43_daynight_spectra} for a comparison between an equilibrium and a quenched model]{cooper06,agundez14,drummond18}.


\section{Bright Star Secondary Eclipse Program} \label{sec:bright}

\subsection{Scientific motivation: resolving atmospheric thermal structures and energy budgets}
Atmospheric thermal structures (i.e., how temperature varies with altitude/pressure) are a crucial diagnostic of how planets absorb and re-radiate the energy they receive from their host stars (i.e., their ``energy budgets''). Theory and observations suggest that the thermal structures of close-in planets may be fundamentally different from those of similar temperature brown dwarfs or young, directly-imaged planets, which are heated from below \citep[e.g.,][]{seager05,burrows06,fortney08}. In particular, thermal inversions due to absorption of short-wavelength radiation at high altitudes may be common in exoplanets, as they are for the Solar System planets \citep{hubeny03}. Recently there have been new indications of thermal inversions in the hottest close-in planets \citep{haynes15,evans.2017.ugews,sheppard17}. However, the existing constraints in this area are poor because the limited wavelength coverage and resolution of current data probe only a narrow range of pressures, capture only a small fraction of the total emitted energy, and don't assay the full range of chemical species that play a role in the energy budget. Thermal emission measurements obtained at secondary eclipse using \textit{JWST} will lead to a dramatic advance in our ability to determine and understand the diverse thermal structures of externally-irradiated exoplanets by resolving many spectral features at high resolution and characterizing the full energy budgets of the planets. Specifically for our fiducial target WASP-18b, we will be able to robustly test for the presence of opacity sources (e.g., H$^-$), and physical processes (e.g., molecular dissociation) occurring in the atmosphere as proposed by \cite{arcangeli_h-_2018}.

\subsection{Technical motivation: bright star limits}
As part of our ERS program we will observe a single secondary eclipse of a hot Jupiter orbiting a bright host star using NIRISS/SOSS. This observation will not only demonstrate the utility of \textit{JWST} data for revealing the atmospheric thermal structures and energy budgets of transiting exoplanets, it will also enable us to determine how precisely \textit{JWST}'s instruments can measure transit spectra in the limit of low photon noise (i.e., a high number of recorded photoelectrons). The Bright Star Program, by pushing the expected noise to very low levels, will test \textit{JWST}'s behavior at the limit of its achievable precision, in preparation for follow up of the compelling transiting exoplanets around bright stars that \textit{TESS} will find. The performance of \textit{JWST} in this regard is unknown, as there are no design requirements, yet it is a key metric that will ultimately determine if terrestrial exoplanet atmospheres are accessible. 

The data obtained from the Bright Star Program will reveal how the instrument systematics change with the fluence, which is something that has been observed for \textit{Hubble}/WFC3 \citep{berta12,wilkins14}. The results of this test are applicable to all the NIR instruments and all planet types, as the dominant systematics in the bright star regime are expected to be related across the common HgCdTe detectors and readout electronics. These data will also be the most representative of typical follow-up observations of bright \textit{TESS} targets in future cycles. Given that \textit{TESS} is expected to deliver hundreds of exciting Earths and super-Earths around bright stars, it is essential that the community develops the tools for the analysis of bright stars early in the mission. Furthermore, bright stars maximize our ability to understand systematic effects that only become apparent at high precision. Evaluating the limits of precision attainable with \textit{JWST} will be essential in Cycle 2 and beyond when assessing the feasibility of ambitious programs designed to detect, for example, the compact atmospheres of terrestrial planets.

\subsection{Planned observations}
The nominal target for the Bright Star Program is the hot Jupiter WASP-18b \citep[K\,=\,8.1,][]{hellier09}, while the back-up targets are WASP-38b \citep[K\,=\,7.5,][]{barros11} and KELT-7b \citep[K\,=\,8.0,][]{bieryla15}. WASP-18b has a transit duration of 2.2\,hours (see Table \ref{tab:planets} for target system properties). Thus with the required burn in and baseline (see \S\ref{sec:overview}) plus overhead, we need 8.7 hours to observe this object. 
\added{This NIRISS/SOSS observation is similar to that described for WASP-79b (see Section \ref{sec:planned_trans}), with the following exceptions.
Because of the brightness of the host stars targeted in this program, we have to use the SUBSTRIP96 NIRISS subarray mode (96x2048 pixels) to avoid saturating our K=8.1 primary target. This mode will limit the wavelength coverage slightly compared to the NIRISS/SOSS observations in the Panchromatic Transmission Program, for which we can use a larger subarray that captures the full spectrum (short wavelength cutoff of 0.85 vs.\ 0.6\,$\mu$m).  With 4 groups per integration (11.1 sec) and 2,172 integrations per exposure, we will achieve a total exposure time of 6.68 hours at $\sim$80\% of saturation (5\% non-linearity) for the most illuminated pixels.  The observation efficiency is 80\%.  For target acquisition, we will achieve a SNR of $\sim$72 using 19 groups in the SOSS Bright mode.}

\subsection{Expected results}
\begin{figure}
\begin{center}
\includegraphics[width=0.45\textwidth]{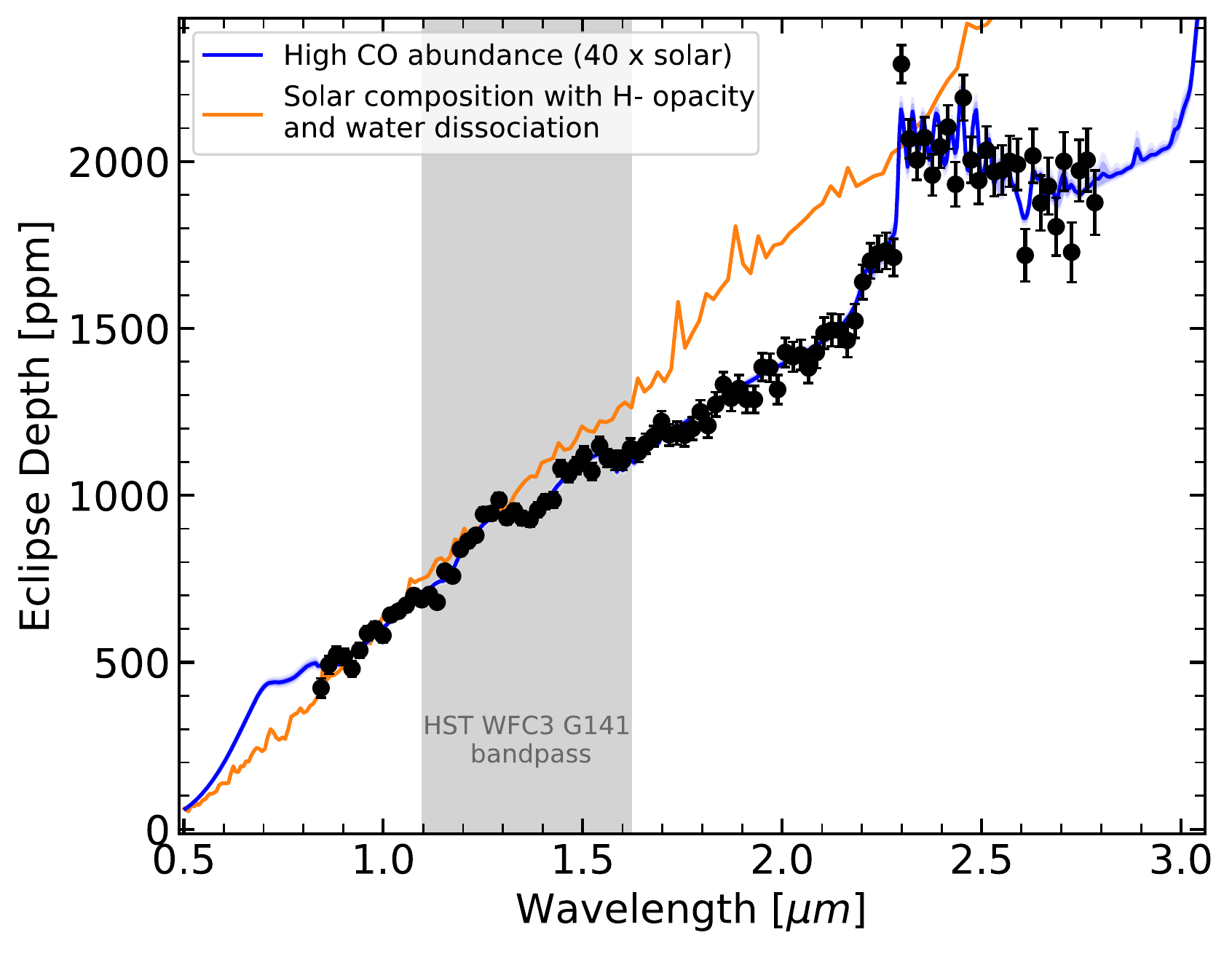}
\caption{\label{fig:WASP18b_Spectra} Modeled thermal emission spectra for our bright star target WASP-18b  (continuous colored curves), compared to simulated \textit{JWST}/NIRISS secondary eclipse observations (points with 1\,$\sigma$ uncertainties). The models are the competing scenarios for explaining the existing \textit{Hubble}/WFC3 data from \citet[][blue line]{sheppard17} and \citet[][orange line]{arcangeli_h-_2018}. The models are degenerate in the WFC3 bandpass given the precision of the existing data but will be easily distinguished with the NIRISS observations}.
\end{center}
\end{figure}

The emission spectrum obtained over the 0.85 - 2.8\,$\mu$m bandpass in the Bright Star Program will capture 80\% of the total thermal emission of WASP-18b (shown in Figure~\ref{fig:WASP18b_Spectra} as eclipse depth versus wavelength). The coverage and fidelity of the energy budget and temperature structure of this highly-irradiated exoplanet will be unprecedented. In contrast, \textit{Hubble}/WFC3 captures $<$20\% of the thermal emission from similarly hot planets (and much less for cooler planets), and it does so with much lower precision and resolution than \textit{JWST}. Importantly, emission spectroscopy is much less sensitive to clouds than transmission spectroscopy and it probes deeper layers of the atmosphere.

\begin{figure}
\begin{center}
\includegraphics[width=0.40\textwidth]{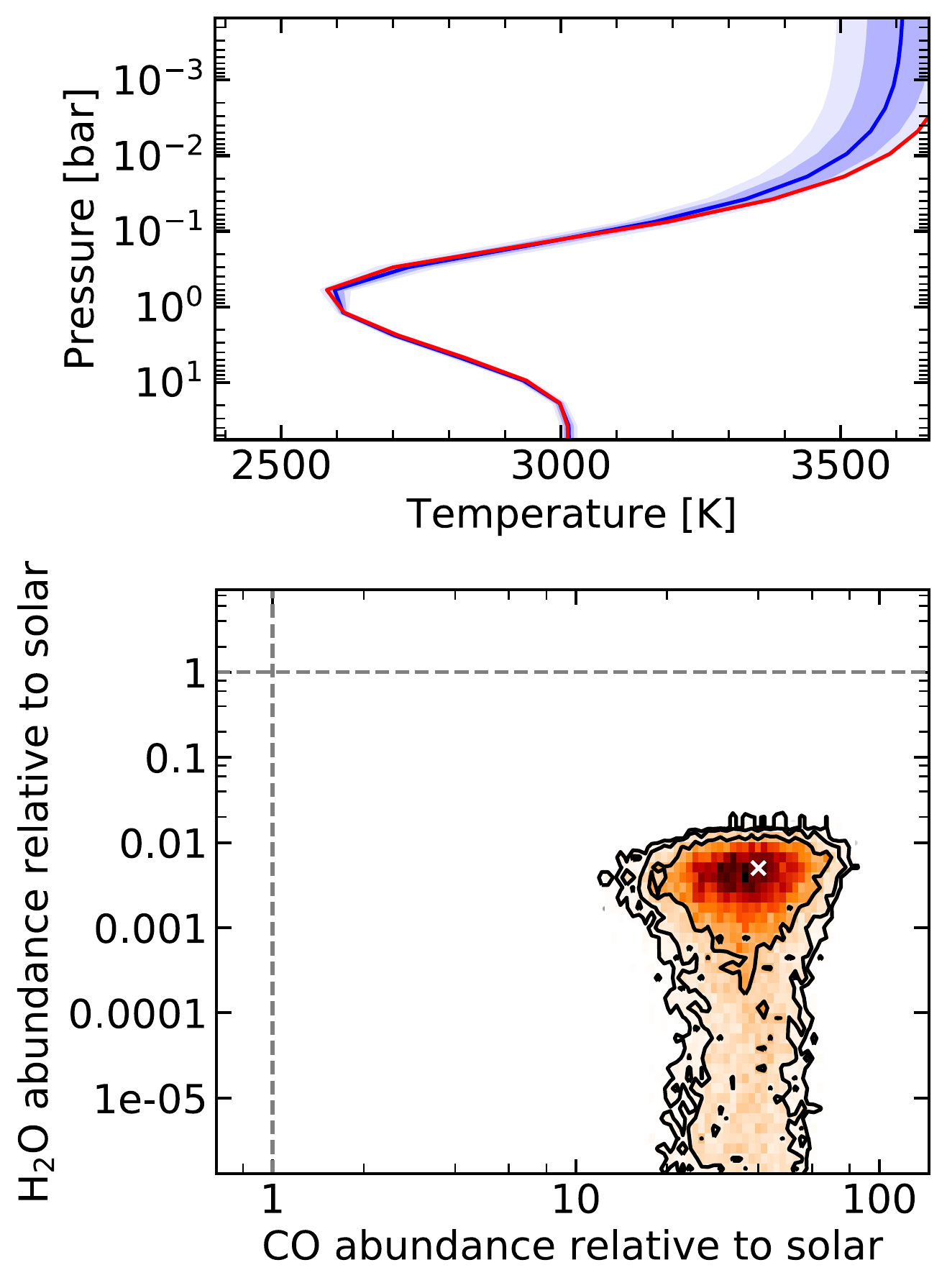}
\caption{Retrieval results for the simulated WASP-18b eclipse measurements with NIRISS. The top panel shows the retrieved temperature profile (blue shading) compared to temperature profile used to simulate the observations (red). The bottom panel shows the 2D-marginalized posterior distribution for the CO and H$_2$O abundances relative to the equilibrium abundances expected for a solar composition gas. The white cross indicates the abundances used to simulate the observations. Here we assumed the high CO, low H$_2$O scenario from \citet{sheppard17} as an example (i.e., a model with no H$^-$ opacity and a constant water abundance with altitude, which is the model represented by the blue line in Figure~\ref{fig:WASP18b_Spectra}). The NIRISS data will clearly yield precise and accurate constraints on the thermal structure and composition of the planet's atmosphere. \label{fig:WASP18b_retrieval}}
\end{center}
\end{figure}

We estimated the NIRISS/SOSS noise for one eclipse of the nominal target WASP-18b using \texttt{PandExo} assuming the best-fit atmospheric model to existing \textit{Hubble}/WFC3 data for this object from \citet{sheppard17}. \added{\texttt{PandExo} accounts for photon noise as well as detector read noise and dark current, whereas photon noise vastly dominates the noise budget of our bright star observations. Systematic drifts due to instrument settling or telescope jitter are not formally considered in our analysis; instead, we assume that these effects can be sufficiently detrended similar to the analysis of \textit{HST}/WFC3 observations. We conservatively inflate the transit depth uncertainties by 10-20\% of the photon-noise limit as regularly achieved \textit{HST}/WFC3 observations. We believe this to be realistic because detector technology of the NIRISS instrument is similar to \textit{HST}/WFC3 and the NIRISS/SOSS mode spreads the light over many pixels in the cross-dispersion direction similar to \textit{HST}/WFC3 spatial scan mode. In addition, unlike \textit{HST} observations, the \textit{JWST} observations provide the advantage of uninterrupted observation of the complete eclipse and the baseline, which should significantly simplify the detrending process. Our planned observations include two hours of baseline before and after the transit to achieve the same SNR inside and outside of the transit. Two hours of baseline before and after should also be sufficient to correct for the instrument drift over the course of the transit observation.}

\added{To assess the scientific potential of the observations, we} applied a classic ``free'' retrieval model \citep{benneke2012,benneke15} to simulated observation and extracted the temperature structure and composition of the atmosphere. \added{In the particular scenario discussed here, we first determined a best-fitting molecular composition and temperature structure by fitting the available \textit{HST}/WFC3 eclipse depth measurements of WASP-18b \citep{sheppard17,arcangeli_h-_2018} and then derived the thermal emission spectrum within the range covered by NIRISS. With a dayside brightness temperature of 2900\,K, WASP-18b is substantially hotter than most hot Jupiters. At such high temperatures there are potentially important contributions to the atmospheric opacity from H$^-$ ions, as well as the removal of major molecules by thermal dissociation (including water) and thermal ionization of metals \citep{arcangeli_h-_2018,kreidberg18b,mansfield18,parmentier18,lothringer18,kitzmann18}. These effects can combine to give a very similar spectrum in the WFC3 bandpass as a high metallicity and high C/O model that doesn't include H$^{-}$ opacity and assumes constant molecular abundances with altitude \citep[i.e., the model from][]{sheppard17}. Both scenarios are shown in Figure~\ref{fig:WASP18b_Spectra}, and we simulate the \textit{JWST} observations and perform the retrieval for the latter scenario with high metallicity and without H- opacity (blue curve).}

\added{The results of our retrieval based on the simulated \textit{JWST} observations are shown in Figure~\ref{fig:WASP18b_retrieval}. We expect to measure the temperature structure of hot planetary atmospheres to the unprecedented precision level of a few percent using NIRISS secondary eclipse observations. This precision on the energy budget and resulting thermal structure will yield fundamentally new insight to radiative transfer in planetary atmospheres. In the specific case of WASP-18b, the NIRISS observations in our ERS program will be able to easily distinguish between these models for the planet by definitively revealing the different spectral features that are expected to be present in broad bandpass observations.}


\section{Enabling Transit Science with \textit{JWST}} \label{sec:analysis}

\added{One core goal of this ERS program is to establish roadmaps for robust analyses of \textit{JWST} transiting exoplanet observations, as early as possible in the mission. For each of our six representative datasets, we plan to publish analysis resources for future \textit{JWST} observers, including both (1) a data analysis tutorial and (2) a report on the instrument's performance for time-series observations. Although we expect each dataset will exhibit its own surprising quirks that demand individualized analyses, many components of the necessary analyses can probably be standardized across all the observations. In this section, we outline our plans for these data analysis recipes (\S\ref{sec:recipes}) and instrument performance reports (\S\ref{sec:fieldguides}), focusing primarily on aspects that are likely to be common across all datasets. In \S\ref{sec:community}, we briefly highlight our plans to engage the transiting exoplanet community through a public ERS data challenge.

\begin{table}[b]
\caption{\label{tab:ingredients} 
Core Ingredients for Data Analysis Toolkits}
\begin{tabular}{l|p{0.9\columnwidth}}
    \hline

1& Visualize the time-series cube of 2D images, with static pixel-by-pixel mean and variance images and movies. \\
    \hline
2 & Extract 1D spectra and their predicted uncertainties, using both fixed apertures and optimal extractions. Measure time-series diagnostics that may inform instrumental models below. \\
    \hline
3 & Separate the instrumental and astrophysical signals, using physically-motivated causal models, as well as independent, statistical approaches such as Gaussian Process models and Principal/Independent Component Analysis techniques. Establish priors from our physical knowledge of the instrument.\\
    \hline
4 & Create a parameterized model of the planet feature that was observed (transit, eclipse, phase curve), including free parameters for stellar limb-darkening and stellar variability. Establish priors from our knowledge of the exoplanet system. \\
    \hline
5 & Fit this joint model to data, using MCMC or nested sampling to estimate the parameters' posterior probability distribution. \\
    \hline
6 & Extract the planetary spectra that are embedded within those fits, after marginalizing over all other parameters and possible instrumental models. These spectra constitute the core scientific measurements of the program, to be archived on MAST. \\
    \hline
    
\end{tabular}
\end{table}

\begin{table*}[htp]
\caption{\label{tab:fieldguide} 
Basic Outline of Instrument Performance Reports}
\begin{tabular}{p{0.45\textwidth}|p{0.45\textwidth}}
    \hline

{\bf The field guides will measure diagnostics...	}
& {\bf ...to help answer basic questions about the instrument. }\\
    \hline

$\bullet$ the number of photons detected per wavelength	 	
& Are \texttt{PandExo}/\texttt{Pandeia}'s core throughput estimates and instrument models accurate? \\
    \hline

$\bullet$ the measured variance of the flux residuals compared to photon noise predictions
$\bullet$ tests for non-Gaussianity of the flux residuals	 	
& Is the spectrophotometry photon-limited, or are there other significant time-series noise sources? \\
    \hline

$\bullet$ the measured variance of time-binned flux residuals vs. temporal bin size
$\bullet$ the power spectrum and autocorrelation function of the flux residuals
& Is the noise correlated in time? How will this limit \textit{JWST}'s precision for exoplanet observables? \\
    \hline

$\bullet$ the position/width/shape of the spectral trace vs. time 
$\bullet$ the background level and reference pixel values vs. time
& How stable are the telescope/instrument optics and detectors over hours-to-days timescales? \\
    \hline

$\bullet$ the strength and form of correlations between the residuals and other available time series (the above image diagnostics, temperature sensors, telescope pointing, antenna movements)	
& What physically-motivated models might explain systematic noise in time-series measurements? \\
    \hline

$\bullet$ the measured variance of wavelength-binned residuals vs. wavelength bin size
$\bullet$ a matrix of correlation strength between all possible wavelength bins
& What instrumental systematics are ``common-mode''? How well can we separate overlapping wavelengths? \\
    \hline

$\bullet$ the descriptive morphology of any other time-dependent trends in the measured spectrophotometry 
& (for example) What is the timescale of detector persistence/charge-trapping? How long does \textit{JWST} need to settle at the start of an observation? \\
        \hline

\end{tabular}
\end{table*}

\subsection{Data Analysis Tutorials}\label{sec:recipes}

Numerous calculations stand between raw pixel readouts initially downloaded from \textit{JWST} and final measurements of planetary transmission spectra, emission spectra, or phase curve properties. We plan to publish data analysis tutorials centered around each of our datasets, to serve as a reference for future \textit{JWST} observers. Table \ref{tab:ingredients} summarizes the main ingredients we plan to include in these data analysis recipes.

The \textit{JWST} Data Reduction Pipeline will handle processing of individual integrations into calibrated 2D slope images \citep[][]{stsci.2016.jdrp}. For each time-series observation (TSO), the \textit{JWST} pipeline will also perform ensemble processing to produce a time series of extracted stellar spectra for the entire observation \citep[see][]{stsci.2016.jto}. Such special treatment of time-series data represents a new feature that was not included in the \textit{Hubble} pipelines, and we plan to fully validate the steps going into the \textit{JWST} TSO pipeline through independent tests. When working toward the limit of extreme precision, experience has shown that big variations can emerge from seemingly small decisions in the extraction process, such as how cosmic rays are mitigated \citep{zhang.2018.ntstpscmts}, how centroids are calculated \citep{agol.2010.c1fftems}, how wavelength shifts are estimated \citep{deming13}, or how extraction apertures are defined \citep{croll.2015.ntednjsgnp}.

We will explore a suite of tools for modeling and mitigating instrumental systematics (see Figure \ref{fig:systematics}). Physically-motivated causal models can provide insight to the processes that contribute to instrumental systematic noise sources, such as telescope motion combined with intra- or interpixel sensitivity variations \citep{ballard.2010.sscnmcwsio, christiansen.2011.spttseceshttwfnemo} or charge-trapping in detector pixels \citep{zhou.2017.pmccthstwfcndatebd}. In some cases, these physical models of the instrument can be approximated through analytic functions or low-order polynomial expansions of other measured parameters \citep[][]{brown01, charbonneau.2008.biese1, agol.2010.c1fftems, burke.2010.notjx, knutson12, deming.2015.ssedmgehupd, luger.2016.epldlc}. These approximations can sometimes suffer from being too rigid in their assumptions, but marginalizing over multiple families of systematics models has been shown to improve their robustness \citep{gibson.2014.rielpudssm, wakeford16}. Since all physical models will inevitably be imperfect descriptions of the instrument, we will also employ more flexible noise-modeling frameworks, including Gaussian Process models \citep{gibson12,danielski13,evans15,cloutier17,sedaghati17,foreman-mackey.2017.fsgpmwaats}, Principal Component Analysis \citep{foremanmackey15,thatte10,zellem14}, and Independent Component Analysis \citep{waldmann12,waldmann14,morello15,morello16,damiano17}. We will compare independent analyses with these different methods both to ensure our scientific results are robust and to provide guidance to the community on the strengths of each method in the \textit{JWST} context.

Standard tools already exist to model the relevant exoplanet signals, such as the Python packages \texttt{batman} \citep{kreidberg.2015.bbtmcp} and \texttt{spiderman} \citep{louden18}. However, \textit{JWST}'s new level of precision demands we pay a new level of attention to several astrophysical signals that can potentially contaminate the inferred exoplanet spectra. For example, the use of fixed, inaccurate limb-darkening coefficients and/or orbital parameters may impart trends on the derived planetary transmission spectra \citep{csizmadia13, espinoza15, parviainen.2015.lldt, morello17}. \textit{JWST's} precision will permit direct tests of limb-darkening models at intermediate resolution \cite[following work by][at lower resolution]{knutson.2007.uslrp2,knutson.2011.stse4esvcdfv}. Likewise, starspots, plages, and other inhomogeneities on the unocculted portion of the stellar disk can introduce spurious transit-depth variations with wavelength, which might mask a planet's real transmission spectrum \citep{berta.2011.gsssvtsap,rackham.2018.tlsefsfidmtp,zhang.2018.ntstpscmts}. We will ensure that our final planetary spectra account for and marginalize over such astrophysical systematics.

\

\subsection{Time-Series Instrument Performance Reports}\label{sec:fieldguides}

The community can already predict \textit{JWST}'s expected photon-limited noise for transit observations, either directly from the \texttt{pandeia} exposure time calculator \citep{pontoppidan.2016.pmetcjw, stsci.2016.jetco} or from its transit-optimized wrapper \texttt{PandExo} \citep{batalha.2017.pctteswj}. However, the only way to know how closely we will be able to approach this predicted instrumental noise performance is to analyze real on-sky data. For each of these ERS observations, we plan to calculate a suite of metrics to assess systematic noise sources and collect the lessons learned into performance reports describing each instrument's capabilities for precise time-series spectrophotometry. These field guides to instrumental systematics will cover the general diagnostics included in Table \ref{tab:fieldguide}, as well as other more instrument-specific issues as discussed in the technical motivations above. Together, these diagnostics aim to test the hypothesis that the instrument, its calibration pipeline, and systematics modeling can collectively result in close-to-ideal photon-counting measurements. They focus particularly on understanding temporal and wavelength correlations in the data, because they have significant potential to corrupt statistical inferences of planet properties \citep{pont.2006.enptd, carter.2009.peftdwcewmatlc,cubillos.2017.caaelc}. These diagnostics can hopefully help inform noise models for careful statistical inference of planet properties and serve as initial inputs for constructing physically-motivated systematics mitigation models.



\subsection{Community Engagement}\label{sec:community}
Another core goal of our ERS program is to catalyze broad engagement in \textit{JWST} and to train a community of capable \textit{JWST} exoplanet observers. To address this goal, we will host a multi-phase data challenge to spark worldwide collaboration and focus the exoplanet community's creativity on analyzing \textit{JWST} data. Inspired by the \textit{Spitzer} 2015 Data Challenge \citep{ingalls16}, this challenge will comprise online interaction and face-to-face meetings, bringing together instrument/telescope specialists, observers, and theorists. It will facilitate the speedy validation of our scientific results and construction of our science-enabling products, through intermediate deadlines and opportunities for group work. These activities are not limited to those scientists who were on the original ERS proposal; we welcome participation from the entire community. }


\section{Conclusion} \label{sec:conclusion}

The multitude of recently discovered transiting exoplanets presents both challenges and opportunities. The \textit{challenge} is to understand these objects as part of a complete theory of planetary system cosmogony, which is one of the preeminent topics of modern astrophysics and planetary science. The \textit{opportunity} is the chance to study a diverse and large sample of planets, including Solar System analogues in different physical regimes (e.g., hot Jupiters and potentially habitable planets around M dwarfs) and classes of planets with no Solar System counterparts (e.g., super-Earths). Now that we have tight constraints on the occurrence rate of these planets \cite[e.g.,][]{fulton.2017.csirdsp}, the next frontier is to obtain a comprehensive census of their atmospheres.

\textit{JWST} holds the promise of enabling a comprehensive census of transiting exoplanet atmospheres that will yield a dramatic advance in our understanding of planetary nature, origins, climate, atmospheric physics and chemistry, and habitability. It is difficult to overstate just how much \textit{JWST} will likely advance this field given its increased capabilities compared to existing facilities. The community has lofty goals \citep[e.g.,][]{barstow16,crossfield17,louie18,kempton18}, and with the recent development and approval of the Transiting Exoplanet ERS Program we are on track to achieve these ambitions.

\acknowledgments
We thank David Charbonneau for strategic advice during the development of our ERS proposal and the anonymous referee for suggestions that improved this paper. M.R.L.\ acknowledges support from the NASA Exoplanet Research Program award NNX17AB56G. J.M.D.\ acknowledges support from the European Research Council (ERC) under the European Union's Horizon 2020 research and innovation programme (grant agreement No.\ 679633, Exo-Atmos). L.D.\ acknowledges support from the Fund of Scientific Research Flanders. J.L.\ acknowledges that this project has received funding from the European Research Council (ERC) under the European Union's Horizon 2020 research and innovation programme (grant agreement No.\ 679030, WHIPLASH). L.M.\ acknowledges support from the Italian Minister of Instruction, University and Research (MIUR) through FFABR 2017 fund and from the University of Rome Tor Vergata through ``Mission: Sustainability 2016'' fund. I.P.W.\ acknowledges funding from the European Research Council (ERC) under the European Union’s Horizon 2020 research and innovation programme (grant agreement No.\ 758892, ExoAI). Part of the research was carried out at the Jet Propulsion Laboratory, California Institute of Technology, under contract with the National Aeronautics and Space Administration.

\end{document}